\documentclass[twocolumn, 10pt]{article}

\usepackage{2023ConferencePHMSociety}
\usepackage{graphicx}
\usepackage{amsmath}

\PassOptionsToPackage{hyphens}{url}
\usepackage{url}
\usepackage{booktabs}

\def\writing{1}
\if\writing1\newcommand{\w}[2][]{\textcolor{red}{\textbf{\ifx&#1&\else\textbf{\begin{large}$\star$\end{large}#1:}\fi} \textit{#2}}}\else\newcommand{\w}[2][]{}\fi

\begin{document}

\title{Sound-Dr: Reliable Sound Dataset and Baseline Artificial Intelligence System for Respiratory Illnesses}

\author{%
	Truong V.~Hoang\authorNumber{1}
	\and 	Quang H.~Nguyen\authorNumber{2}
        \and 	Cuong Q.~Nguyen\authorNumber{3}
        \and 	Phong X.~Nguyen\authorNumber{4}
	\And 	Hoang D.~Nguyen\authorNumber{5}
}

\address{
	\affiliation{{1,3,4}}{AI Center, FPT Software Company Limited, Vietnam}{ 
		{\email{\{truonghv1, cuongnq1, phongnx1\}@fsoft.com.vn}}
		} 
	\affiliation{2}{Reliable Machine Learning Group, Vietnam}{ 
		{\email{nh.quang313@gmail.com}}} 
	\affiliation{5}{School of Computer Science and Information Technology, University College Cork, Ireland}
		{\email{hn@cs.ucc.ie}}
}

\maketitle
\pagestyle{fancy}
\thispagestyle{plain}

\phmLicenseFootnote{Truong V.~Hoang}

\begin{abstract}
As the burden of respiratory diseases continues to fall on society worldwide, this paper proposes a high-quality and reliable dataset of human sounds for studying respiratory illnesses, including pneumonia and COVID-19. It consists of coughing, mouth breathing, and nose breathing sounds together with metadata on related clinical characteristics. We also develop a proof-of-concept system for establishing baselines and benchmarking against multiple datasets, such as Coswara and COUGHVID. Our comprehensive experiments show that the Sound-Dr dataset has richer features, better performance, and is more robust to dataset shifts in various machine learning tasks. It is promising for a wide range of real-time applications on mobile devices. The proposed dataset and system will serve as practical tools to support healthcare professionals in diagnosing respiratory disorders. 
The dataset and code are publicly available here: https://github.com/ReML-AI/Sound-Dr/.
\end{abstract}

\section{Introduction}
\label{sec1}

Abnormalities can be discovered in the respiratory sounds of individuals with fever, asthma, tuberculosis, pneumonia, and COVID-19 compared to the sound of those without these conditions. A solid body of literature has shown the effectiveness of respiratory sounds in disease detection with the use of artificial intelligence (AI) \cite{proven_2, proven_4, parkinson}. 
Furthermore, AI systems and data can be periodically updated, thereby improving accuracy and reliability. In real-world situations, sound-based medical screening tools can be widely deployed in multiple locations, such as airports, factories and supermarkets.

To date, there are several respiratory sound datasets to detect diseases, such as the Internal Conference on Biomedical Health Informatics (ICBHI) data \cite{ICBHI} in which each audio recording identifies the patients in terms of being healthy or exhibiting one of the following respiratory diseases or conditions including COPD, Bronchiectasis, Asthma, Upper and Lower respiratory tract infection, Pneumonia, and Bronchiolitis. To detect COVID-19, there are also two well-known datasets from New York \cite{newyork} and Cambridge \cite{cambridge} universities. These respiratory datasets, however, are likely prone to reliability issues, as shown in our later experiments with the use of a dataset shift detection method \cite{loudly}.

In order to build a high-quality and reliable dataset, we developed a system to collect respiratory sound data in an efficient manner, such as recording each sound multiple times to reduce the impact of unwanted noises and capture the average sample's duration longer for better reliability. As a result,  our dataset, named Sound-Dr, is collected under many different diseases, such as fever, asthma, and COVID-19, to enable researchers to solve various machine-learning problems related to respiratory diseases, including disease classification and anomaly detection. 

The Sound-Dr dataset contains three types of respiration sounds, including nose breathing, mouth breathing, and coughing, with extensive lengths to foster more possibilities in machine learning algorithms and model deployments. Besides the audio recordings, metadata and health-related characteristics (e.g., smoking, insomnia) are included with high quality and data richness, which can be useful for multiple machine learning tasks on medical diseases related to respiratory systems.
Compared to existing datasets, the Sound-Dr dataset has multiple advantages with the following contributions:
\begin{itemize}
    \item In collaboration with medical experts \cite{woolcock}, we contribute a publicly available, high-quality, and reliable dataset with a sampling rate of 48,000 Hz and an average duration of 23s for respiration sounds, including breathing (nose \& mouth) and coughing. 
    \item Besides the audio recordings, the dataset also provides metadata and health-related characteristics for various tasks in machine learning, including but not limited to disease classification, anomaly detection, and symptom recognition for respiratory illnesses. It is suitable for large-scale adoption and deployment via smart devices for homes or businesses. 
    \item This paper establishes an open baseline framework to facilitate benchmarking the Sound-Dr dataset and other datasets in terms of performance and robustness, as well as the efficiency of the data collection.
\end{itemize}

\section{Background}
\label{sec:background}
\subsection{Studies of Human Sounds for Medical Screening}

In medicine, human sounds have been well-studied as viable inputs for identifying vocal fold pathology, which involves either subjective or objective assessments. In subjective approaches, a skilled medical professional hears the sound signal and determines whether it is diseased or normal based on their prior training and experience. Nevertheless, depending on the level of experience, this type of evaluation may differ from doctor to doctor \cite{kreiman1990listener}. As a result, both medical and engineering professionals are paying more and more attention to objective approaches to voice pathology. 

Many medical conditions can be accurately identified using computer-aided voice pathology classification tools and deep learning techniques. 
For example, a recent study \cite{related_1} proposed a deep learning-based classification model that can accurately predict whether a person has a cold or not based on their speech on URTIC dataset. This dataset is highly imbalanced, with only 2876 samples having Cold class. On the other hand, No Cold class contains 25776 samples.
Moreover, Mo et al. \cite{related_5} proposes a computer-based deep learning algorithm that could enable rapid screening of the most common pulmonary diseases (COPD, Asthma, and respiratory infection (COVID-19)) using voluntary cough sounds alone. 

During the global pandemic, Sharma et al. \cite{related_6} presented a challenge aimed at accelerating the research in acoustics-based detection of COVID-19, a valuable topic at the intersection of acoustics, signal processing, machine learning, and healthcare. This was an open call with great interest from the research community.

\subsection{Machine learning datasets for respiratory diseases}
The use of machine learning has become increasingly promising for the detection and monitoring of respiratory illnesses. 
A recent work \cite{related_7} presented an exploration of various deep-learning models for detecting respiratory anomalies from auditory recordings. Authors used the ICBHI 2017 \cite{ICBHI}, each audio recording contains one or different types of respiratory cycles, labeled as Crackle, Wheeze, Both Crackle \& Wheeze, or Normal. Using a late fusion of inception based and transfer learning frameworks outperforms state of the art, recording the best score of 57.3\%. 
Another study \cite{related_2} applied a CNN for discriminating pathological voices from normal voices with the speech samples from Saarbrücken Voice Database, which is a collection of speech and electroglottography signals of more than 2000 speakers. Resulting in F1 scores of 78.7\%. 

There are also recent datasets for respiratory diseases, such as COUGHVID \cite{coughvid} and Coswara \cite{coswara}. COUGHVID is a global cough signal recordings dataset for COVID-19 detection with some clinical information and metadata. And Coswara is another dataset composed of voice samples from healthy individuals, including breathing sounds (fast and slow), cough sounds (deep and shallow), phonation of sustained vowels, and counting numbers. 
These datasets are large-scale and regularly updated; nevertheless, they are susceptible to reliability issues due to their intrinsic properties and distributional characteristics.
We use these two datasets for performance benchmarking and evaluate the dataset shift problem.

\section{Sound-Dr Dataset and Task Definition}
\label{sec:dataset}
\subsection{Sound-Dr Dataset Collection}
Sound-Dr dataset is a project
\footnote{\url{https://www.fpt-software.com/fpt-softwares-ai-app-listens-out-for-respiratory-diseases-amid-looming-covid-19}}
of AI Center of FPT Software Company Limited \cite{fptsoftware}, an entity in charge of AI research and development with consultation from the Woolcock Institute of Medical Research, Vietnam~\cite{woolcock}. The application collects users’ demographic information, medical history, and records of voices and respiratory sounds. The Sound-Dr dataset was solely collected by FPT for community purposes during the peak season of the COVID-19 pandemic in Vietnam from August 2021 to October 2021. We treat ethical issues as important, and users were prompted to read about our terms and conditions and give us their consent solely for development and community purposes. Therefore, the dataset has the agreement and consent of all users.

In the data collection, we developed web-based and mobile-based applications for users to easily interact and record three different sounds: (1) mouth breathing, (2) nose breathing, and (3) coughing. With the involvement of medical experts in the field, for each audio type, users are requested to record at least three times with a minimum duration of 5 seconds in each turn. The sample rate of 48,000 Hz is set to be the default, and no noise reduction method is used in web-based or mobile-based applications to collect the true nature of the data. Additionally, some metadata of users is also collected via a survey form which includes personal information (e.g., age and gender), related respiratory illness symptoms, smoking status, and COVID-19 diagnosis, as shown in Table \ref{table:metadata}.

\begin{table}[!ht]
  \begin{center}
  \caption{Metadata fields of the Sound-Dr dataset.}
  \label{table:metadata}
  \setlength{\tabcolsep}{2pt}
  \scalebox{0.8}{
  \begin{tabular}{p{0.8in}p{1.1in}p{2.1in}}
    \toprule
    \textbf{Categories} & \textbf{Fields} & \textbf{Details} \\
    \midrule
        Demographics
            & Sex options & Gender: Male, Female \\
            \cmidrule(r){2-3} & Ages & Age \\
            \cmidrule(r){2-3} & Current city & The current city living \\
        \cmidrule(r){1-3}
        Related-Flu Symptoms & Symptoms status choice & Symptoms since last 14 days: Fever, Chills, Sore throat, Dry cough, Wet cough, Stuffy nose, Snivel, Headache, Difficulty breathing or feeling short of breath, Muscle aches, Dizziness, Confusion or vertigo, Tightness in your chest, Loss of taste and smell, None \\
        \cmidrule(r){1-3}
        Medical Conditions & Condition choice & The medical conditions of the subject: Asthma, Diabetes, Cystic fibrosis, COPD/Emphysema, Pulmonary fibrosis, Other lung diseases, Angina, Previous stroke or Transient ischaemic attack, Cancer, Previous heart attack, Valvular heart disease, HIV or impaired immune system, Other long-term condition, Other heart diseases, Previous organ transplant, None \\
        \cmidrule(r){1-3}
        Insomnia Symptoms & Insomnia status choice & How often the subject suffers from insomnia: Never, Once in the last 2 weeks, Once a week, 2 to 3 days a week, 4 days a week or more \\
        \cmidrule(r){1-3}
        Smoking
            & Smoking status & Never smoked, Ex-smoker, \\
            Status & options & Current smoker (less than once a day), Current smoker (1-5 cigarettes per day), Current smoker (11-20 cigarettes per day), Current smoker (21+ cigarettes per day),  \\
        \cmidrule(r){1-3}
        COVID-19 Status
            & Cov19 status choice & How long has had a positive test for COVID-19: Never, In the last 14 days, More than 14 days ago.\\
            \cmidrule(r){2-3} & F condition choice & Status with COVID-19 of a subject \\
    \bottomrule
  \end{tabular}
  }
  \end{center}
  \vspace{-0.3cm}
\end{table}

\subsection{Descriptive Statistics of Sound-Dr Dataset}

\begin{figure*}[!ht]
    \begin{minipage}{0.45\linewidth}
        	\centering
        	\includegraphics[scale=0.28]{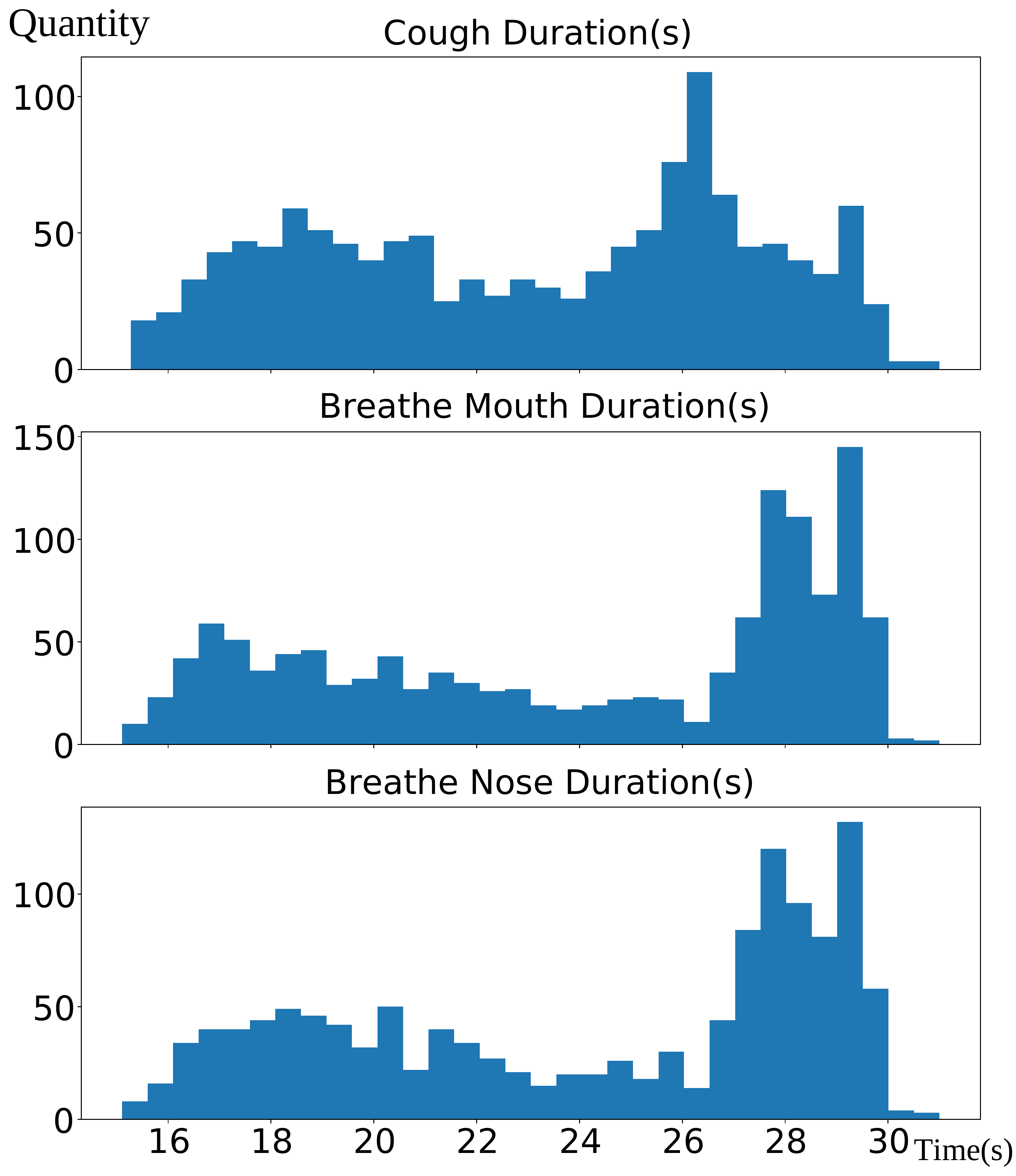}
            \caption{Histograms of three types of audio recording duration.}
            \label{fig:duration}
    \end{minipage}\hfill
    \begin{minipage}{0.45\textwidth}
    	    \centering
    	    \includegraphics[width=0.6\linewidth]{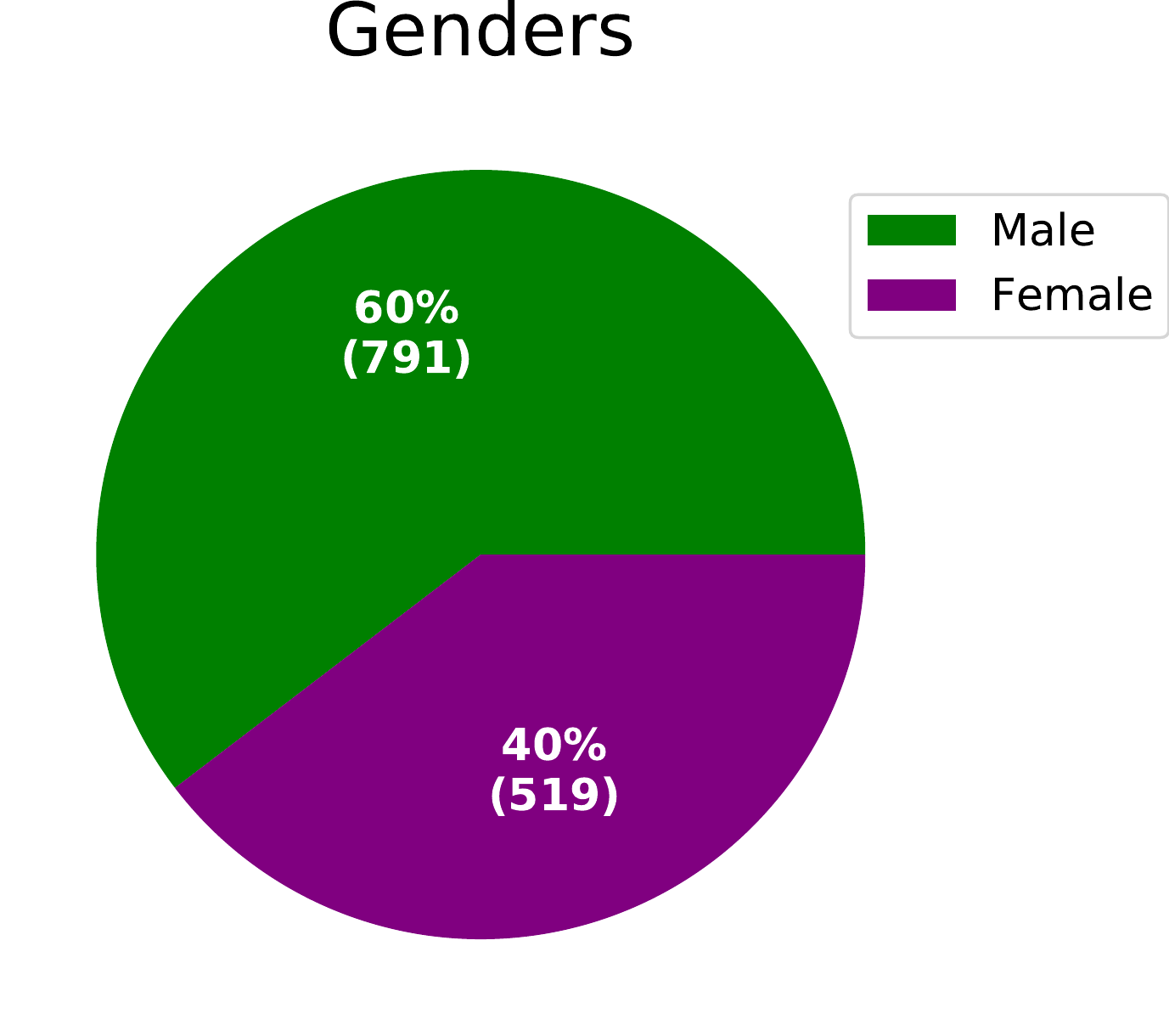}
            \caption{Gender Data Distribution.}
            \label{fig:gender}
    	    \centering
    	    \includegraphics[width=0.9\linewidth]{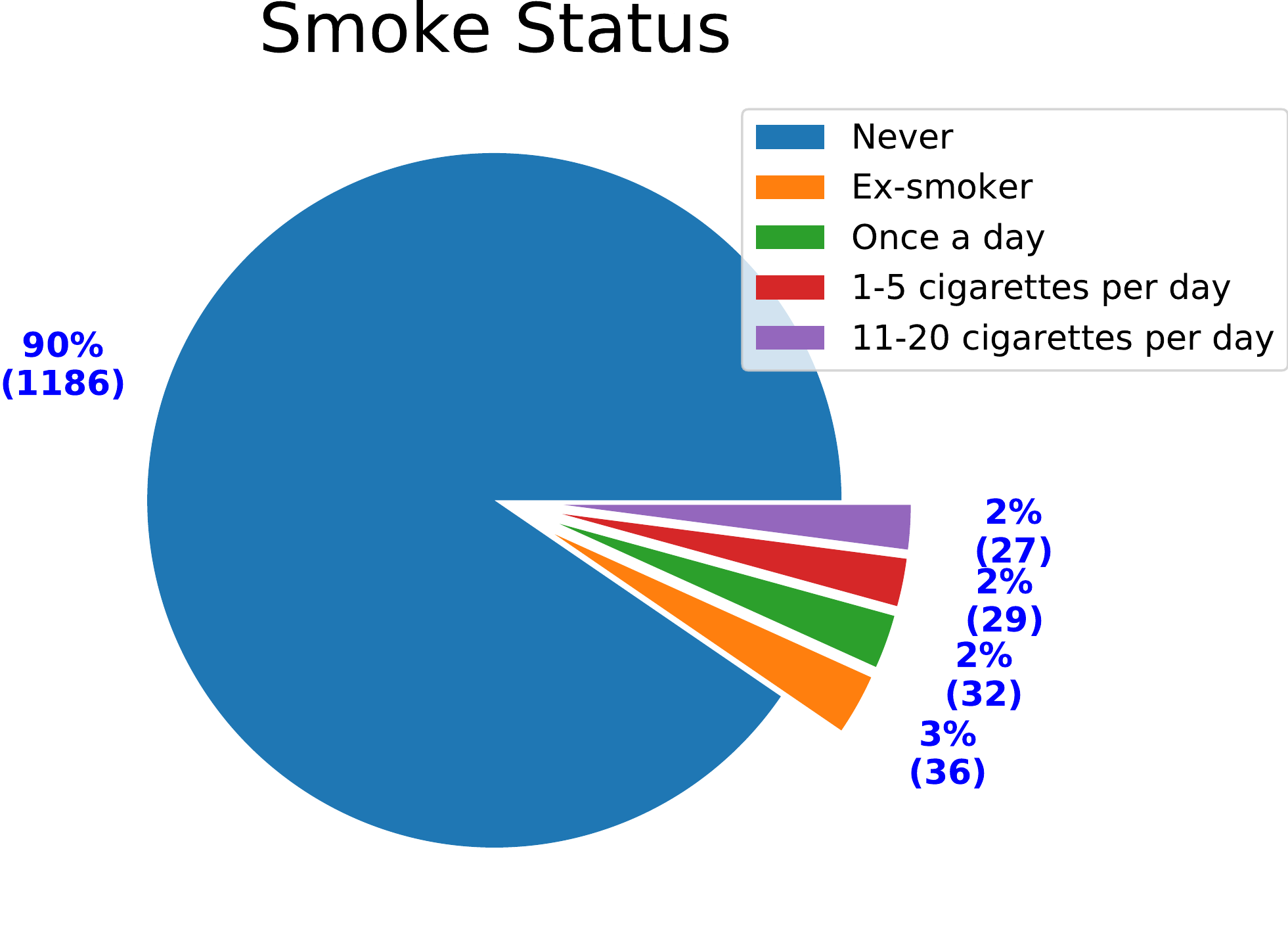}
            \caption{Smoke Status Distribution.}
            \label{fig:smoke}
    \end{minipage}
\end{figure*}

\begin{figure}[ht]
    \centering
    \includegraphics[width=0.9\linewidth, height=1.5in]{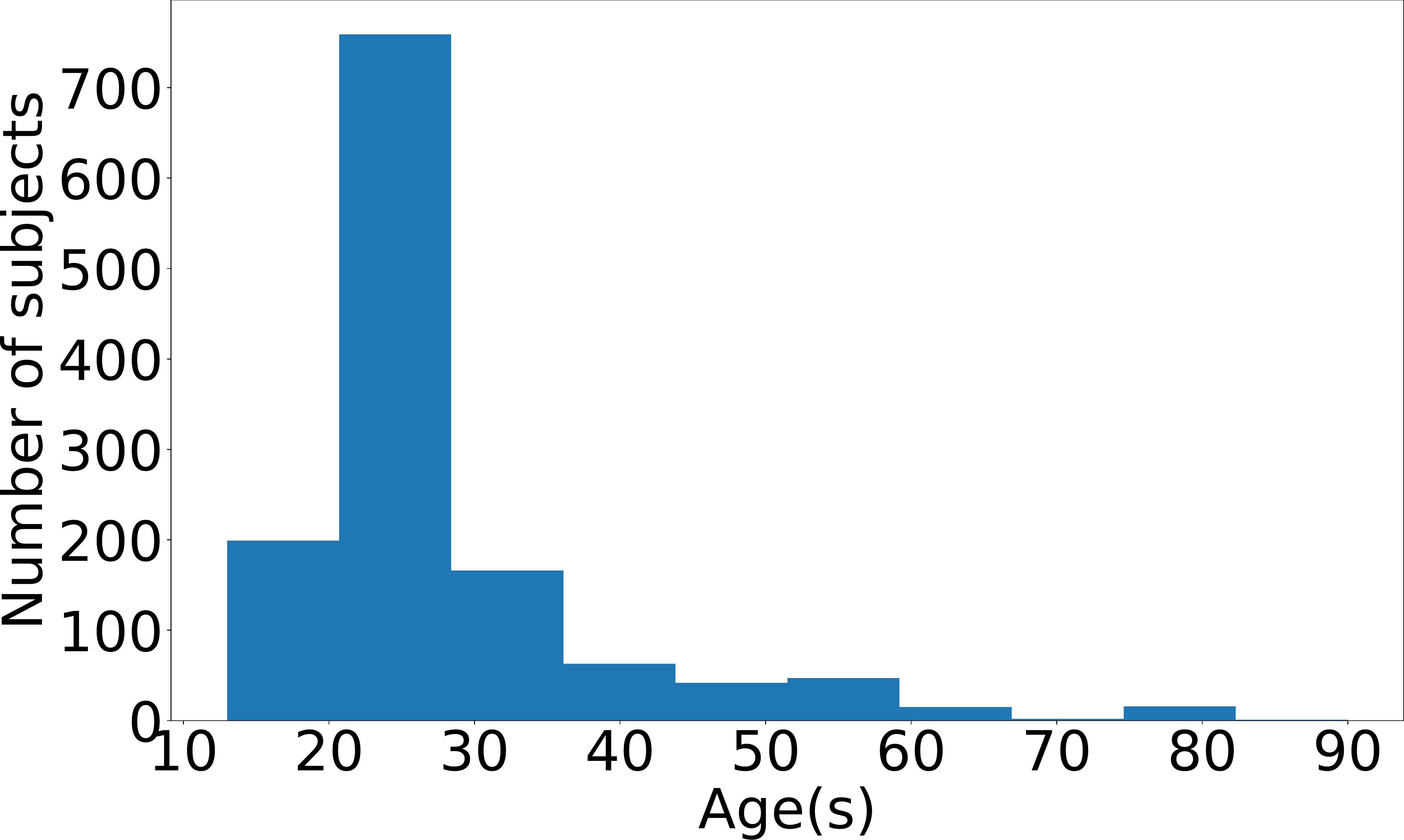}
    \caption{Age Group Data Distribution.}
    \label{fig:age}
\end{figure}

We obtained a dataset of 3,930 sound recordings; the distribution of coughing, mouth breathing, and nose breathing is presented in Figure~\ref{fig:duration}. There are 1,310 in total subjects with gender distribution shown in Figure \ref{fig:gender} and age distribution presented in Figure \ref{fig:age}. It can be seen that more males (e.g. 60\%) than females (e.g. 40\%) participated in our program. In terms of age groups, subjects between 20 and 40 years old are dominant. Regarding the smoking status, Figure~\ref{fig:smoke} indicates that 90\% of the subjects are non-smokers.
From 346 COVID-19 positive subjects, there are 293 objects with symptoms and 193 objects without symptoms.

\begin{figure}[ht]
    \centering
    \includegraphics[width=0.9\linewidth, height=1.5in]{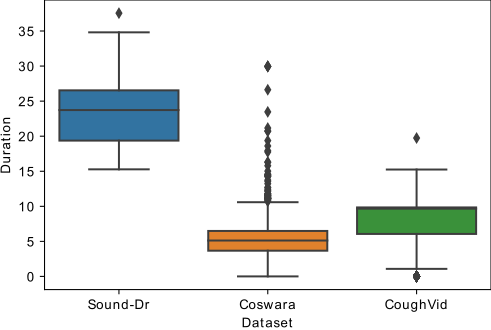}
    \caption{The distribution of the duration of datasets.}
    \label{fig:dataset_duration}
\end{figure}

\begin{figure}[ht]
    \centering
	\includegraphics[width=1\linewidth]{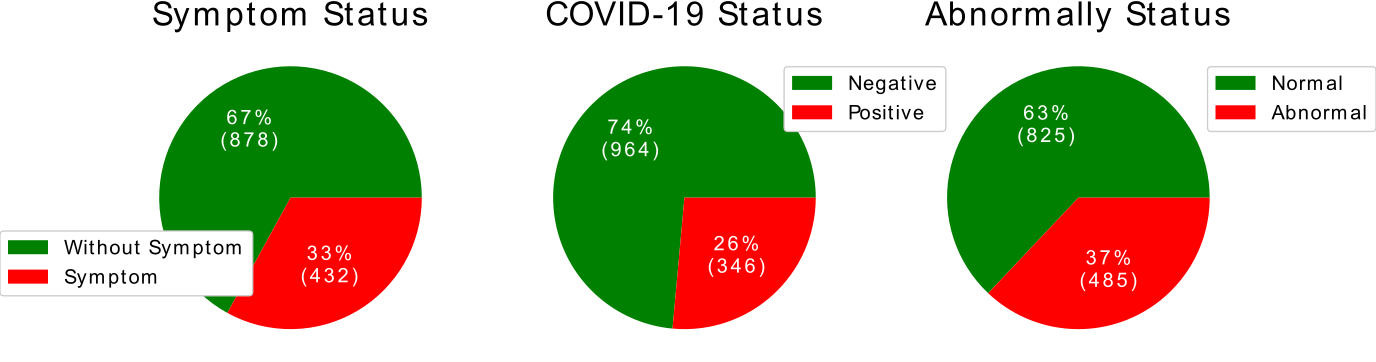}
    \caption{The number of subjects for each task defined.}
    \label{fig:covid_num}
\end{figure}

Statistics of the duration, shown in Figure \ref{fig:dataset_duration}, reveal several interesting characteristics in quantifying above mentioned datasets. Some lengths of audio in Coswara and COUGHVID samples are less than 1s, which might lead to being unqualified for the training model and errors in the reading input data process. Nevertheless, the Sound-Dr dataset has longer durations to ensure that training data is split into even parts and does not require padding when the sound length is unsatisfactory. Moreover, regarding statistics of the sampling rate, the Coswara and COUGHVID datasets have audio sampling rates of 44100Hz and 22050Hz, respectively. The Sound-Dr dataset has a higher sampling rate; therefore, our training data can be easily converted to different sampling rates for benchmarking. Our data collection system was built skillfully to ensure the best quality for machine learning tasks.

\subsection{Task Definition}
\label{task_def}
Given the Sound-Dr dataset, we propose three main tasks: (I) Detect negative or positive COVID-19 subjects, (II) Detect subjects with and without related respiratory symptoms, and (III) Detect normal subjects and anomaly subjects (i.e., anomaly subjects are positive COVID-19 or present related respiratory symptoms). For each task, the audio input of coughing, mouth breathing mouth, and nose breathing are evaluated independently.

Based on the metadata as shown in Table \ref{table:metadata}, the total of 1,310 subjects are separated into: COVID-19 negative and COVID-19 positive subjects, subjects with and without symptoms, and normal and anomaly subjects for task I, II, and III respectively as shown in Figure~\ref{fig:covid_num}.

To evaluate the Sound-Dr dataset for each defined task, we apply a 5-fold cross-validation method where the final result is an average of all folds. Random seeds are used to ensure that results are reproducible and the data is divided the same way into all methods for benchmarking. Every experiment's result is an average of the 5 different seeds. The evaluation metrics in use are Accuracy (Acc), F1 score~\cite{f1}, and AUC~\cite{auc}. All the model training, benchmarking, and evaluation tasks have been executed in a system with Ubuntu 18.04, 12 GB RAM, and NVIDIA GTX 1080 GPU.

\section{Baseline System}
\label{sec:model}
Given the Sound-Dr dataset, we develop a deep-learning-based framework to explore, which is referred to as the baseline. Generally, the baseline framework can be separated into two main steps: Feature extraction and Classification.

\subsection{Feature Extraction}
The raw audio from one channel (mono) is firstly re-sampled with a sample rate of 16000 Hz using Librosa toolkit \cite{librosa}. 
Then, re-sampled audio recordings are fed into a pre-trained model to extract embedding features. In this paper, the pre-trained model is from both TRILL~\cite{trill} and FRILL~\cite{frill}, which is recommended for downstream tasks on non-semantic speech signals. Using TRILL to extract features from Cough sounds for detecting COVID-19 has also been proven effective \cite{top2_seconddicova}.

\begin{table}[ht]
  \small  
  \begin{center}
  \caption{The setting parameters of Classification}
  \label{table:setting_backend}
  \setlength{\tabcolsep}{3pt}
  \scalebox{0.9}{
  \begin{tabular}{p{1.0in}p{0.5in}p{1.9in}}
    \toprule
    \textbf{Tasks} & \textbf{Models} & \textbf{Setting Parameters} \\
    \midrule
        \textbf{Task I}: COVID-19 Subject Detection & XGB
            & max\_depth = 6, learning\_rate = 0.07, scale\_pos\_weight = 2.78, n\_estimators = 200, subsample = 1, colsample\_bytree = 1, eta = 1, objective = `binary:logistic', eval\_metric = `auc' \\
        \cmidrule(r){1-3}
        \textbf{Task II}: Symptom Subject Detection, \textbf{Task III:} Anomaly Subject Detection & XGB, XGBOD
            & max\_depth = 7, learning\_rate = 0.3, scale\_pos\_weight = 1.7, n\_estimators = 200, subsample = 1, colsample\_bytree = 1, nthread = -1, eval\_metric = `logloss' \\
            \cmidrule(r){2-3}
            & Isolation Forest & n\_estimators = 500, max\_samples = 'auto', contamination = 0.1 \\
    \bottomrule
  \end{tabular}
  }
  \end{center}
\end{table}

While the pre-trained TRILL model is based on ResNet architecture presenting a large footprint, the pre-trained FRILL model is built on MobileNet architecture, leveraging knowledge distillation from the pre-trained TRILL model. As a result, the pre-trained FRILL model is suitable for real-time application on edge devices (i.e., FRILL pre-trained model is 32 times faster on a Pixel 1 smartphone and equals to 40\% of TRILL size, but still competitive to TRILL model with an average decrease of only 2\% in terms of accuracy).

The outputs of both pre-trained models are a time series of one embedding. This means we obtain one embedding (i.e., 2048-dimensional vector) from every second when feeding the audio recordings with different lengths from the Sound-Dr dataset into the models. Hence, we obtain multiple embeddings representing one audio recording. Consequently, we conduct two statistical features of mean and standard deviation across the time axis. We then concatenate these features to create the final embedding (4096-dimensional vector).

\subsection{Classification}

We conduct experiments on the Support Vector Machine, Random Forest, Multilayer Perceptron, ExtraTrees Classifier, LightGBM, and XGB Classifier. Anomaly detection mostly focuses on unsupervised or semi-supervised settings, we use Isolation Forest \cite{isolation_forest}, and XGBOD \cite{XGBOD} for actually seeing the usage of this dataset for anomaly detection for recognizing the outliers. However, we only achieved a good score on XGB Classifier for both Coswara and COUGHVID. Therefore, to build a baseline system and classify extracted embedding features into certain groups defined in Section \ref{task_def}, we use XGB Classifier \cite{xgboost_classify}. To fine-tune hyper-parameters of this classifier, shown in Table \ref{table:setting_backend}, we make use of the Optuna framework \cite{optuna_2019} with the Grid Search algorithm. All these classification models are implemented by using XGBoost library \cite{xgboost} for XGB Classifier, Python Outlier Detection library \cite{pyod} for XGBOD, and Scikit-Learn toolkit \cite{scikit-learn} for the others.

\begin{table}[ht]
  \small  
  \begin{center}
  \caption{The experimented results on Sound-Dr dataset over five runs. Results in \textbf{bold font} mark the best results given the same (fair) task.}
  \label{table:results}
  \setlength{\tabcolsep}{2pt}
  \scalebox{0.9}{
  \begin{tabular}{p{0.5in}p{1.0in}p{0.6in}p{0.3in}p{0.3in}c}
    \toprule
    \textbf{Data} & \textbf{Feature - Classifier} & \textbf{Detection} & \textbf{Acc} & \textbf{F1} & \textbf{AUC} \\ Type & & Task & \textbf{Mean} & \textbf{Mean} & \textbf{Mean (Std)} \\
    \midrule
        Cough & TRILL - XGB & Symptom & 78.78 & 67.22 & 80.86 (00.58) \\
         \cmidrule(r){3-6}
         & & COVID-19 & \textbf{86.56} & 71.34 & 86.56 (01.09) \\
         \cmidrule(r){3-6}
         & & Abnormal & \textbf{77.25} & 67.46 & 79.20 (00.66) \\
         \cmidrule(r){2-6}
         & FRILL - XGB & Symptom & \textbf{79.05} & \textbf{67.67} & \textbf{81.23 (00.33)} \\
         \cmidrule(r){3-6}
         & & COVID-19 & 86.06 & \textbf{73.13} & \textbf{88.44 (00.38)} \\
         \cmidrule(r){3-6}
         & & Abnormal & 77.18 & \textbf{68.12} & \textbf{81.16 (01.11)} \\
        \cmidrule(r){1-6}
        Breathing & TRILL - XGB & Symptom & 74.43 & 62.65 & 78.45 (01.03) \\
         \cmidrule(r){3-6}
         Mouth & & COVID-19 & 82.75 & 67.44 & 85.55 (00.79) \\
         \cmidrule(r){3-6}
         & & Abnormal & 75.64 & 65.66 & 78.03 (01.28) \\
         \cmidrule(r){2-6}
         & FRILL - XGB & Symptom & 79.31 & 65.65 & 80.57 (00.65) \\
         \cmidrule(r){3-6}
         & & COVID-19 & 86.18 & 70.76 & 87.23 (01.14) \\
         \cmidrule(r){3-6}
         & & Abnormal & 75.04 & 66.04 & 78.79 (00.62) \\
        \cmidrule(r){1-6}
        Breathing & TRILL - XGB & Symptom & 78.47 & 63.75 & 78.75 (01.08) \\
         \cmidrule(r){3-6}
         Nose & & COVID-19 & 86.11 & 70.16 & 85.04 (00.72) \\
         \cmidrule(r){3-6}
         & & Abnormal & 76.34 & 64.37 & 78.24 (00.71) \\
         \cmidrule(r){2-6}
         & FRILL - XGB & Symptom & 79.54 & 65.19 & 79.80 (00.86) \\
         \cmidrule(r){3-6}
         & & COVID-19 & 84.50 & 70.28 & 85.79 (00.97) \\
         \cmidrule(r){3-6}
         & & Abnormal & 77.10 & 67.74 & 80.75 (00.93) \\
    \bottomrule
    \multicolumn{6}{l}{
        \begin{tabular}[c]{@{}l@{}}
        \small{Abnormal: Symptom + COVID-19.} \\
        \end{tabular}
      }
  \end{tabular}
  }
  \vspace{-0.5cm}
  \end{center}
\end{table}

\subsection{Experimental Results and Discussion}
\label{subsec:result}
We experimented with the task of COVID-19 Detection based on the three collected sound types: Cough, Breathing mouth, and Breathing nose, as illustrated in Table \ref{table:results}. The performance using Breathe Mouth and Breathe Nose is lower compared to the Cough sound data. The best performance using Cough sound scores 88.44 AUC, 73.13 F1, 86.06 Accuracy. Although TRILL outperforms FRILL on Accuracy by about 0.2\% (86.06-86.26 Acc), on F1 and AUC metrics, FRILL performs better for 2\% (73.13-71.34, 88.44-86.56 AUC). Therefore, we use FRILL for our baseline model as it is satisfactory for the real environment that needs fast, accurate detection, especially on mobile devices.

In addition, we also experiment with the Abnormal Detection in respiratory sound by adjusting the label which we combine the COVID-19 Positive and Symptomatic status into Abnormal labels. Using XGB Classifier with hyper-parameters shown in Table \ref{table:setting_backend}, we achieve promising results of 81.16 AUC, 68.12 F1, and 77.18 Accuracy. On XGBOD, results of 82.95 AUC, 70.02 F1, and 79.77 Accuracy show that the unsupervised settings can be used on this dataset for anomaly detection. The performance comparison is described in Table \ref{table:benchmark_unsupervised_results}. This shows that our dataset has potential for more reliable outcomes on multiple tasks, such as Outlier Detection and Anomaly Detection in Respiration Sound. We hope that models based on the Sound-Dr dataset could be built to support the doctor's diagnosis of disease faster and more accurately in the future.

\begin{table}[!ht]
  \small  
  \begin{center}
  \caption{The benchmark results of unsupervised methods on other datasets over five runs for Abnormal Detection task (Symptom + COVID-19). Results in \textbf{bold font} mark the best results given the same (fair) dataset.}
  \label{table:benchmark_unsupervised_results}
  \setlength{\tabcolsep}{2pt}
  \begin{tabular}{p{0.65in}p{1.2in}ccc}
    \toprule
    \textbf{Data Type} & \textbf{Feature - Classifier} & \textbf{Acc} & \textbf{F1} & \textbf{AUC} \\ & & \textbf{Mean} & \textbf{Mean} & \textbf{Mean (Std)} \\
    \midrule
        Coswara & FRILL - IsolationForest & 76.63 & 16.37 & 49.82 (02.99) \\
        & FRILL - XGBOD & 76.23 & 41.47 & 67.44 (01.25) \\
        \cmidrule(r){1-5}
        COUGHVID & FRILL - IsolationForest & 74.99 & 15.38 & 49.45 (00.68) \\
        & FRILL - XGBOD & 49.62 & 33.96 & 58.84 (01.62) \\
        \cmidrule(r){1-5}
        \textbf{Sound-Dr} & FRILL - IsolationForest & 60.31 & 15.86 & 53.64 (00.54) \\
        \textbf{(Ours)} & FRILL - XGBOD & \textbf{79.77} & \textbf{70.02} & \textbf{82.95 (01.33)} \\
    \bottomrule
  \end{tabular}
  \end{center}
\end{table}

\section{Benchmarks}
\label{sec:benchmarks}

\subsection{Dataset Shift Detection}

Besides evaluating the effectiveness of the models applied on 3 datasets, we parallelly consider the dataset shift problem that contributes to measuring the dataset's robustness. It happens due to the different distributional characteristics of data between train and test set \cite{dataset_shift}. 

\begin{table}[!ht]
  \small  
  \begin{center}
  \caption{Detecting Dataset Shift Using Failing Loudly}
  \label{table:loudly}
  \begin{tabular}{p{1.1in}ccccc}
    \toprule
    \textbf{Data type} & \multicolumn{5}{c}{\textbf{Number of samples from test }}  \\
    \cmidrule(r){2-6} & 50 & 100 & 500 & 1000 & 10000 \\
    \midrule
        Coswara & 0.28 & 0.44 & 0.43 & 0.43 & 0.42 \\
        COUGHVID & 0.39 & 0.44 & 0.40 & 0.80 & 0.41 \\
        \textbf{Sound-Dr (Ours)} & \textbf{0.25} & \textbf{0.29} & \textbf{0.39} & \textbf{0.38} & \textbf{0.38} \\
    \bottomrule
  \end{tabular}
  \end{center}
  \vspace{-0.5cm}
\end{table}

\begin{table}[!ht]
  \small  
  \begin{center}
  \caption{The benchmark results of supervised methods on other datasets over five runs. Results in \textbf{bold font} mark the best results given the same (fair) feature.}
  \label{table:benchmark_results}
  \setlength{\tabcolsep}{2pt}
  \scalebox{0.85}{
  \begin{tabular}{p{0.65in}p{1.1in}p{0.6in}p{0.3in}p{0.3in}c}
    \toprule
    \textbf{Data Type} & \textbf{Feature - Classifier} & \textbf{Detection Task} & \textbf{Acc} & \textbf{F1} & \textbf{AUC} \\ & & & \textbf{Mean} & \textbf{Mean} & \textbf{Mean (Std)} \\
    \midrule
        Coswara & FRILL - SVM & COVID-19 & 74.97 & 40.94 & 65.12 (01.42) \\
                         \cmidrule(r){3-6}
                         & & Abnormal & 70.78 & 39.16 & 58.71 (00.92) \\
                         \cmidrule(r){2-6}
                         & OpenSmile - SVM & COVID-19 & 74.83 & 43.58 & 67.61 (01.69) \\
                         \cmidrule(r){3-6}
                         & & Abnormal & \textbf{68.24} & 32.85 & 58.70 (01.33) \\
                         \cmidrule(r){2-6}
                         & DeepSpectrum - & COVID-19 & \textbf{75.19} & 17.52 & 51.48 (01.71) \\
                         \cmidrule(r){3-6}
                         & SVM & Abnormal & \textbf{72.38} & 19.36 & 51.31 (01.66) \\
                         \cmidrule(r){2-6}
                         & OpenXBoW 1000 - & COVID-19 & 76.92 & 41.07 & 64.97 (01.16) \\
                         \cmidrule(r){3-6}
                         SVM & & Abnormal & \textbf{73.10} & 29.52 & 56.91 (02.10) \\
                         \cmidrule(r){2-6}
                         & OpenXBoW 2000 - & COVID-19 & 80.28 & 42.29 & 65.15 (01.40) \\
                         \cmidrule(r){3-6}
                         SVM & & Abnormal & \textbf{73.74} & 29.51 & 56.99 (01.27) \\
                         \cmidrule(r){2-6}
                         & OpenXBoW 3000 - & COVID-19 & 78.01 & 43.98 & 65.01 (02.21) \\
                         \cmidrule(r){3-6}
                         SVM & & Abnormal & \textbf{76.87} & 30.56 & 56.00 (01.98) \\
        \cmidrule(r){1-6}
        COUGHVID & FRILL - SVM & COVID-19 & 81.49 & 16.89 & 54.60 (00.63) \\
                         \cmidrule(r){3-6}
                         & & Abnormal & 64.69 & 23.55 & 50.89 (00.43) \\
                         \cmidrule(r){2-6}
                         & OpenSmile - SVM & COVID-19 & \textbf{80.80} & 15.68 & 53.66 (01.35) \\
                         \cmidrule(r){3-6}
                         & & Abnormal & 65.34 & 25.01 & 51.89 (01.12) \\
                         \cmidrule(r){2-6}
                         & DeepSpectrum - & COVID-19 & 49.78 & 14.15 & 49.02 (01.32) \\
                         \cmidrule(r){3-6}
                         & SVM & Abnormal & 49.00 & 27.55 & 49.53 (00.58) \\
                         \cmidrule(r){2-6}
                         & OpenXBoW 1000 - & COVID-19 & \textbf{79.47} & 15.77 & 53.67 (01.87) \\
                         \cmidrule(r){3-6}
                         SVM & & Abnormal & 72.12 & 21.62 & 52.26 (01.40) \\
                         \cmidrule(r){2-6}
                         & OpenXBoW 2000 - & COVID-19 & \textbf{86.20} & 11.45 & 51.86 (01.67) \\
                         \cmidrule(r){3-6}
                         SVM & & Abnormal & 69.84 & 21.40 & 51.37 (00.94) \\
                         \cmidrule(r){2-6}
                         & OpenXBoW 3000 - & COVID-19 & \textbf{83.04} & 15.00 & 55.35 (00.84) \\
                         \cmidrule(r){3-6}
                         SVM & & Abnormal & 67.67 & 21.59 & 50.72 (00.78) \\
        \cmidrule(r){1-6}
        \textbf{Sound-Dr} & FRILL - SVM & COVID-19 & \textbf{82.53} & \textbf{70.48} & \textbf{81.37 (00.85)} \\
             \cmidrule(r){3-6}
             \textbf{(Ours)} & & Abnormal & \textbf{76.11} & \textbf{69.45} & \textbf{75.54 (00.26)} \\
             \cmidrule(r){2-6}
             & OpenSmile - SVM & COVID-19 & 69.31 & \textbf{56.96} & \textbf{71.75 (01.41)} \\
             \cmidrule(r){3-6}
             & & Abnormal & 65.04 & \textbf{59.33} & \textbf{65.83 (01.92)} \\
             \cmidrule(r){2-6}
             & DeepSpectrum - & COVID-19 & 64.58 & \textbf{38.62} & \textbf{57.41 (01.51)} \\
             \cmidrule(r){3-6}
             & SVM & Abnormal & 58.02 & \textbf{45.22} & \textbf{55.71 (00.95)} \\
             \cmidrule(r){2-6}
             & OpenXBoW 1000 - & COVID-19 & 75.40 & \textbf{59.56} & \textbf{72.77 (01.96)} \\
             \cmidrule(r){3-6}
             SVM & & Abnormal & 64.43 & \textbf{55.19} & \textbf{63.35 (01.65)} \\
             \cmidrule(r){2-6}
             & OpenXBoW 2000 - & COVID-19 & 75.25 & \textbf{60.27} & \textbf{73.42 (02.10)} \\
             \cmidrule(r){3-6}
             SVM & & Abnormal & 66.26 & \textbf{59.60} & \textbf{66.46 (02.00)} \\
             \cmidrule(r){2-6}
             & OpenXBoW 3000 - & COVID-19 & 75.56 & \textbf{59.84} & \textbf{72.99 (01.32)} \\
             \cmidrule(r){3-6}
             SVM & & Abnormal & 68.02 & \textbf{58.31} & \textbf{66.45 (01.69)} \\
    \bottomrule
    \multicolumn{6}{l}{
        \begin{tabular}[c]{@{}l@{}}
        \small{Abnormal: Symptom + COVID-19.}
        \end{tabular}
      }
  \end{tabular}
  }
  \end{center}
  \vspace{-0.5cm}
\end{table}

Many machine learning algorithms are based on the assumption that the training and test data are drawn from the same distribution; thus, dataset shift might lead to the model's tremendous performance degradation. We qualify the robustness of the dataset between Sound-Dr, Coswara, and COUGHVID by detecting accuracy shifts indicating the degree of distribution shifting in the dataset.

We conducted several experiments based on a pipeline for detecting dataset shift by a two-sample-testing-based approach, using pre-trained classifiers for dimensionality reduction \cite{loudly}. Specifically, the train and test set is reduced in dimension and subsequently analyzed via statistical hypothesis testing. We investigate the equivalence of the source distribution (from which training data is sampled) and target distribution (from which real-world data is sampled). The shifting of datasets is evaluated with various amounts of samples including {50,100,500,1000,10000} accordingly. Table \ref{table:loudly} shows that the Sound-Dr dataset exhibits less shifting in train-test distribution. Over samples, our result reached better values of about 15\% and 26\% with respect to Coswara and COUGHVID; thereby leading to lesser risk of drifting and more reliability for real-world deployment.

\subsection{Task Performance}

The paper aims to establish performance benchmarks for multiple machine-learning tasks. We exploit extracted features through SVMs with linear kernels for classification tasks. Specifically, we use several extraction methods including FRILL, OpenSmile \cite{opensmile}, OpenXBOW \cite{openxbow} and Deep Spectrum \cite{deep_spectrum} to extract feature representations from preprocessed raw audio data. Acquired representations were scaled to zero mean and unit standard deviation following the parameters from the respective training set. These normalized features were applied to the SVM model employed by the Scikit-Learn toolkit \cite{scikit-learn} with its class LINEARSVC with the optimized complexity parameter C. We conduct experiments in these settings and unify them to a result in Table \ref{table:benchmark_results}.

We utilise the same feature extraction process and classifier (SVM) for COVID-19 and abnormal detection tasks on datasets. The experiment results on the Sound-Dr dataset are better than the two other datasets in Table \ref{table:benchmark_results}. 
The task performance improvements are statistically significant for both COVID-19 Detection and Abnormal Detection on both the Coswara and COUGHVID datasets respectively.

It demonstrates that the Sound-Dr dataset might provide potential features for detecting anomalies in respiratory sounds such as cough and breath. In addition, better results of the Sound-Dr dataset indicate that our dataset was processed well to obtain high-quality samples during data collection.

\section{Conclusion}
\label{sec:conclusion}

High-quality respiratory sound data, which can be used to detect patient symptoms, is in demand; thus, the Sound-Dr dataset is essential for researchers to build health applications. 
We also build a system to evaluate multiple datasets and create the first baseline system for future research and benchmarking. Based on our comprehensive experimental results, the Sound-Dr dataset is better than multiple existing datasets in terms of both unsupervised and supervised methods. Therefore, the Sound-Dr dataset is effectively collected with extensive lengths to minimize various noises. Furthermore, our dataset's unique properties and metadata of health-related characteristics are more reliable against dataset shifts. 

We build a model using FRILL embedding and XGBoost classifier for potential real-life context that necessitates rapid and accurate detection. It also helps the researchers easy to explore to improve the performance compared with the baselines. With the baseline system and dataset available, researchers have the advantage of rapid development of solutions in high demand. With the Sound-Dr dataset, we hope that researchers accelerate the building of Artificial Intelligence models to support doctors diagnose diseases faster and more accurately. The Sound-Dr dataset is collected from various mobile devices, with rigorous data collection methods, promising to apply widely in real-world situations.

With the increasing impact of respiratory illnesses, the Sound-Dr dataset is proposed in collaboration with medical experts to study respiratory anomalies, including pneumonia and COVID-19. As the baseline, this dataset can be useful to qualify respiratory disease screening/abnormal detection/symptom classification. In real-world scenarios, the dataset has been used in multiple medical apps for rapid screening due to its quality and robustness such as Respiratory diseases, COVID-19, and Respiratory anomalies. 

In our pipeline, more data are needed in the field to enhance neural networks optimally. Although we provide an additional dataset on respiration to increase the distribution of data, more data are needed across many countries with a larger number of subjects. By collecting data from subjects from South East Asia, our research aims to provide the groundwork for future advancements in information processing and machine learning.

\section*{Acknowledgment}
This work is supported by the FPT Software AI Center of FPT Software Company Limited \cite{fptsoftware}. FPT Software is a global technology and IT services provider headquartered in Hanoi, Vietnam.



\begin{singlespace} 

\bibliographystyle{apacite}
\bibliography{main}

\end{singlespace} 
\section*{Biographies}

\begin{wrapfigure}[8]{l}[0pt]{0.7in}
\vspace{-15pt}
\begin{center}
\includegraphics[width=0.83in,height=1in,clip,keepaspectratio]{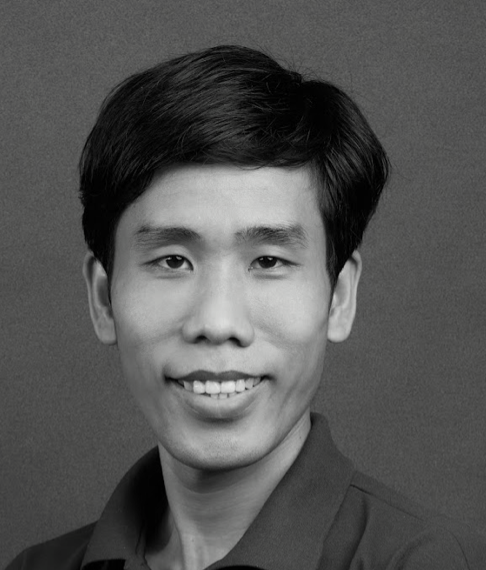}
\end{center}
\end{wrapfigure}
\noindent\textbf{TRUONG V. HOANG} was born in Dong Nai province, Vietnam in 1988. He received B.S. and M.S. degrees in Mathematics in Computer Science from the University of Science, Ho Chi Minh City, in 2016. He achieved Top 2 of The Second DiCOVA Challenge - Track 2: COVID-19 Cough and is a Kaggle Competitions Master. He has been working for more than 5 years as a full-time AI Scientist at the AI Center of FPT Software Company Ltd., HCMC. His main research includes Computer Vision, Audio Processing. Besides, M.S Hoang is also responsible for all technical matters in developing AI products.

\begin{wrapfigure}[8]{l}[0pt]{0.7in}
\vspace{-15pt}
\begin{center}
\includegraphics[width=0.83in,height=1in,clip,keepaspectratio]{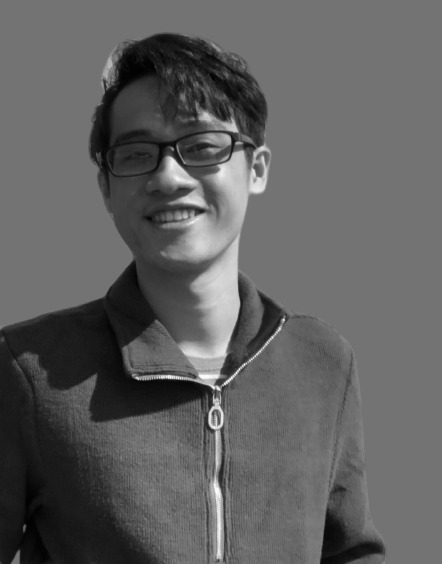}
\end{center}
\end{wrapfigure}
\noindent\textbf{QUANG H. NGUYEN} holds a Master’s
degree in Computer Science at University of Engineering Technology - Vietnam National University, Hanoi in 2022. He has been working in AI Engineer role for more than 3 years. His main research fields include recommendation systems, smart health systems, and reliable artificial intelligence.
He achieved the first prize in the Kalapa Credit Scoring Challenge. Moreover, he participated in the program committee in many competitions, such as Vietnamese Language and Speech Processing, and AI for COVID-19.

\begin{wrapfigure}[8]{l}[0pt]{0.7in}
\vspace{-15pt}
\begin{center}
\includegraphics[width=0.83in,height=1in,clip,keepaspectratio]{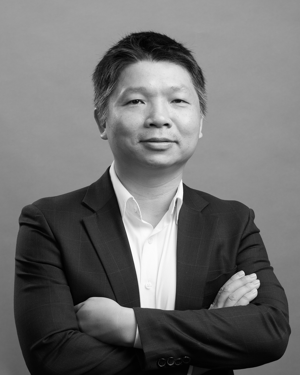}
\end{center}
\end{wrapfigure}
\noindent\textbf{CUONG Q. NGUYEN} is Research Manager with AI Lab, FPT Software Company, Hanoi, Vietnam. He finished his M.Sc. at Andong National University, Sound Korea. His research interests are machine learning and real-time applications of deep learning. He has a long time working on Signal Processing, Computer Vision applications, and real-time OCR. Currently, he managed several applied research projects including sound, computer vision, and numerical.

\begin{wrapfigure}[8]{l}[0pt]{0.7in}
\vspace{-15pt}
\begin{center}
\includegraphics[width=0.83in,height=1in,clip,keepaspectratio]{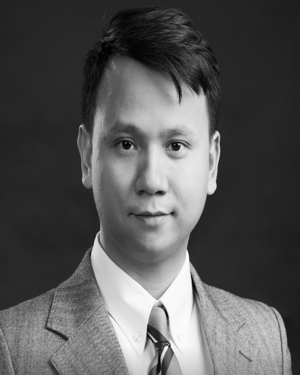}
\end{center}
\end{wrapfigure}
\noindent\textbf{PHONG X. NGUYEN} is currently the Chief AI Officer of FPT Software, where he is playing a leading role in formulating the AI strategy of FPT Software. He sets up many research teams for fundamental and applied research topics like Reinforcement Learning, Efficient Neural Modeling, Representation Learning for Code, sound-based Intelligent Inspection, Computer Vision based OCR, and Image Generation.
 
He received his Ph.D. in Engineering from The University of Tokyo (Japan) and had 7 years of experience working as an Artificial Intelligence researcher for Hitachi, based in Tokyo, Japan. His research focused on Machine Learning for Human Activity Recognition and automatic control for various disciplines. In 2019, he was a Visiting Researcher at the Mila Institute of Artificial Intelligence in Quebec under the direction of Professor Yoshua Bengio - winner of 2018 AM Turing Award. At Mila, Phong researched many projects on Disentanglement, Anomaly Detection, and Reinforcement Learning.

\begin{wrapfigure}[8]{l}[0pt]{0.7in}
\vspace{-15pt}
\begin{center}
\includegraphics[width=0.83in,height=1in,clip,keepaspectratio]{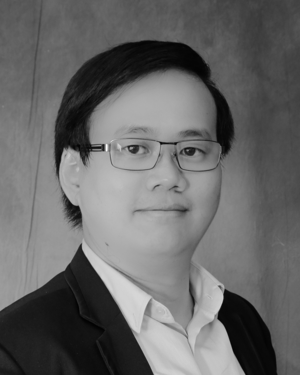}
\end{center}
\end{wrapfigure}
\noindent\textbf{HOANG D. NGUYEN} is currently the Assistant Professor/Lecturer in the School of Computer Science and Information Technology, University College Cork - National University of Ireland, Cork, Ireland. He holds a Ph.D. in Information Systems and Analytics from the National University of Singapore, Singapore. Dr. Nguyen is also affiliated with the SFI Research Centre for Data Analytics (Insight) and SFI Centre for Research Training in Artificial Intelligence (CRT-AI). He has been working on multiple large-scale research projects in reliable artificial intelligence, robust decision optimization and smart health systems. Moreover, Dr. Nguyen has been chairing many AI/ML competitions and events, such as reliable intelligence on social media, AI for COVID-19 detection and MLOps challenges.


\clearpage
\end{document}